\documentclass[twocolumn,showpacs,amsmath,amssymb,prd,nofootinbib]{revtex4}
\usepackage{epsfig}
\def\sss{\scriptscriptstyle}
\def\^#1{^{\sss #1}}
\def\_#1{_{\sss #1}}
\def\beq{\begin{equation}}
\def\eeqno#1{\label{#1}\end{equation}}
\def\ten#1#2{^{\sss#1}_{\sss#2}}

\def\cmss{{\rm cm~s^{-2}}}

\def\az{a\_{0}}
\def\baz{\bar a\_{0}}

\def\l0{\ell\_{0}}
\def\lm{\ell\_{M}}

\def\rar{\rightarrow}
\def\s{\sigma}
\def\l{\lambda}

\def\f{\phi}
\def\fN{\phi\_N}
\def\gfN{\grad\fN}

\def\z{\zeta}

\def\r{\rho}

\def\m{\mu}
\def\n{\nu}

\def\Up{\Upsilon}
\def\C{\Gamma}
\def\A{\mathcal{A}}

\def\F{\mathcal{F}}
\def\K{\mathcal{K}}
\def\L{\mathcal{L}}
\def\U{\mathcal{U}}

\def\Q{\mathcal{Q}}

\def\M{\mathcal{M}}

\def\d{\delta}
\def\drt{d^3r}

\def\a{\alpha}
\def\b{\beta}
\def\c{\gamma}
\def\d{\delta}
\def\eps{\epsilon}
\def\vr{{\bf r}}

\def\vR{{\bf R}}

\def\vF{{\bf F}}

\def\vv{{\bf v}}

\def\va{{\bf a}}

\def\vF{{\bf F}}

\def\vE{{\bf E}}
\def\vB{{\bf B}}

\def\S{\Sigma}

\def\grad{\vec\nabla}

\def\gf{\grad\phi}

\def\gmn{g\_{\m\n}}
\def\Gmn{g\^{\mu \nu}}
\def\gab{g\_{\alpha\beta}}

\def\hGmn{\hat g\^{\mu \nu}}

\def\hgmn{\hat g\_{\m\n}}

\def\hgh{\hat g^{1/2}}
\def\gh{g\^{1/2}}

\def\hC{\hat\C}
\def\cd#1{{}_{\sss;#1}}

\def\oot{\frac{1}{2}}

\def\azg{\A_0}

\def\cmss{{\rm cm~s^{-2}}}

\def\cm{{\rm ~cm}}

\def\gr{{\rm ~gr}}
\def\vfi{\varphi}
\def\EM{E\_M}

\def\fmn{F_{\m\n}}
\def\Fmn{F^{\mu \nu}}

\begin{document}
\title{MOND theory}
\author{Mordehai Milgrom }
\affiliation{Department of Particle Physics and Astrophysics, Weizmann Institute}

\begin{abstract}
A general account of MOND theory is given. I start with the basic tenets of MOND, which posit departure from standard dynamics in the limit of low acceleration  --  below an acceleration constant $\az$ -- where dynamics become scale invariant. I list some of the salient predictions of these tenets. The special role of $\az$ and its significance are then discussed.
In particular, I stress its coincidence with cosmologically relevant accelerations, which may point to MOND having deep interplay with cosmology. The deep-MOND limit and the consequences of its scale invariance are considered in some detail. There are many ways to achieve scale invariance of the equations of motion -- guaranteed if the total action has a well-defined scaling dimension. The mere realization that this is enough to ensure MOND phenomenology opens a wide scope for constructing MOND theories.
\par
General aspects of MOND theories are then described, after which I list briefly presently known theories, both nonrelativistic and relativistic.
With few exceptions, the construction of known, full-fledged theories follows the same  rough pattern: they modify the gravitational action, hinge on $\az$, introduce, already at the level of the action, an interpolating function between the low and high accelerations, and, they obey MOND requirements in the two opposite limits.
These theories have much heuristic value as proofs of various concepts (e.g., that covariant MOND theories can be written with correct gravitational lensing). But, probably, none points to the final MOND theory. At best, they are effective theories of limited applicability.
I argue that we have so far explored only a small corner of the space of possible MOND theories.
\par
I then outline several other promising approaches to constructing MOND theories that strive to obtain MOND as an effective theory from deeper concepts, for example, by modifying inertia and/or gravity as a result of interactions with some omnipresent agent. These have made encouraging progress in various degrees, but have not yet resulted in full-fledged theories that can be applied to all systems and situations.
\par
Some of the presently known theories do enjoy a natural appearance of a cosmological-constant-like contribution that, furthermore, exhibits the observed connection with $\az$. However, none were shown to address fully the mass discrepancies in cosmology and structure formation that are otherwise explained by cosmological dark matter. This may well be due to our present ignorance of the true connections between MOND and cosmology.
\par
We have no clues as to whether and how MOND aspects enter nongravitational phenomena, but I discuss briefly some possibilities.
\end{abstract}
\pacs{04.50.Kd, 95.35.+d, 98.62.Dm}
\maketitle
\section{\label{introduction} Introduction}
MOND is a paradigm of dynamics propounded over thirty years ago \cite{milgrom83} to account for the mass discrepancies in galaxies and galactic systems without dark matter. \footnote{MOND originally stood for ``Modified Newtonian Dynamics'', but has taken on a wider scope and meaning over the years. It was conceived in mid 1981, put forth in three papers in the beginning of 1982, and published in 1983.} It departs from Newtonian dynamics (ND) and general relativity (GR) for very low accelerations, such as are typical of galactic systems. Reference \cite{fm12} is a recent review of MOND.
\par
MOND starts from some ``axioms'' that apply to relativistically-weak-field dynamics [hereafter ``the weak-field limit'' (WFL)], namely, to systems of mass $M$ and size $R$ such that the Newtonian gravitational potentials $\f\sim MG/R\ll c^2$.
\par
1. MOND introduces a constant, $\az$, with the dimensions of acceleration. This constant marks the boundary between the validity domains of the old (ND, GR) and new (MOND) dynamics. In this role it appears saliently in several phenomena that pertain to the transition between the two regimes. $\az$ also appears in many phenomena and regularities that pertain to the very low acceleration regime, far below $\az$ itself.
Such roles are similar to those of $\hbar$ in the quantum/classical context, or to those of the speed of light in the relativty/classical context. In those familiar instances, $\hbar$ or $c$, like $\az$ in MOND, define, by their dimensions, the system attributes that discern the old physics from the new, they also mark the boundary between the classical and the modified regime, and, they appear in many phenomena and regularities in the strongly modified regime.
\par
2. At high accelerations, much above $\az$ -- i.e., when we take the formal limit $\az\rar 0$ in a MOND result or a MOND theory -- standard dynamics is restored. This is analogous to the correspondence principle in quantum theory for the limit $\hbar\rar 0$, or to the nonrelativistic (NR) limit, effected by taking $c\rar \infty$.
\par
3. In the opposite limit of low acceleration $\ll\az$ -- the deep-MOND limit (DML) -- dynamics become space-time scale invariant (SI), namely, invariant under $(t,\vr)\rar\l(t,\vr)$ \cite{milgrom09a}.\footnote{The original description of the DML in Ref. \cite{milgrom83}, and later applications, had posited a relation between the acceleration, $g$, of a test particle and the Newtonian acceleration, $g\_N\sim MG/r^2$, of the form $g\approx (\az g\_N)^{1/2}$. This relation is SI, since under scaling $g\rar\l^{-1}g$, and $g\_N\rar\l^{-2}g\_N$. However, this pristine definition was neither general nor exact, while the definition in terms of SI is both.}
\par
In such a limit, $G$ and $\az$ do not appear separately because of their dimensions, which stymie SI, only the product $\azg\equiv G\az$ appears (see Sec. \ref{dml}).
\par
Two limits in which (gravitational) fields are ``weak'' are relevant to MOND: the WFL, epitomized by $MG/R\ll c^2$, and the DML, epitomized by $MG/R^2\ll\az$; their interplay is discussed e.g., in Ref. \cite{milgrom14b}.
\par
In some contexts, a useful proxy for $\az$ is the MOND length, $\lm\equiv c^2/\az$, and another is the MOND mass $M\_M\equiv c^4/\azg$.
\par
We shall see below that except for the Universe at large, only WFL systems can ever probe the MOND regime. In other words, a strong-field, MOND system must be larger than the Hubble distance. This is why we impose the basic tenets only on the WFL of a MOND theory.
\par
From the above basic tenets follow already many strong predictions \cite{milgrom14}; in fact, practically all the many predictions that have been tested so far. This is very reassuring on one hand. On the other hand, it means that it is difficult to discriminate between MOND theories.
\par
We have today several MOND theories, as well as ideas for constructing more, both NR and relativistic. These have proven very helpful, but, for reasons listed below, in my view we are still a far cry from a fully satisfactory MOND theory.
\par
We do not know to what extent and how MOND affects nongravitational phenomena such as electromagnetism (EM). For example, if there is a consistent way to extend and apply the basic tenets to nongravitational physics.
\par
In Sec. \ref{salient}, I explain some basics of MOND phenomenology, with emphasis on aspects that are particularly pertinent to the construction of theories. Section \ref{azero} points to a possibly special significance of $\az$, its possible connection with cosmology, and some implications of this special value that have important ramifications for MOND theories. Section \ref{dml} treats, in some detail, the crucial DML, with its SI.
In Sec. \ref{structure}, I review general aspects of MOND theory.  Section \ref{theories} starts with a description of the present status of MOND theory, explaining, in particular, why the quest for a fundamental MOND theory is by no means over. It then lists and reviews briefly the presently known MOND theories. Section \ref{ideas} discusses additional theoretical ideas, with some promise for obtaining MOND phenomenology, which, however, have not yet led to complete theories. Section \ref{cosmology} reflects on MOND's roots in, and connections with, cosmology. In Sec. \ref{dmlem}, I comment on possible involvements of MOND in nongravitational phenomena.

\section{\label{salient}Some salient MOND predictions}
Many major predictions of MOND regarding dynamics of galactic systems follow already from the basic tenets listed above. These are discussed thoroughly in Ref. \cite{milgrom14}. This observation is very important, because it shows that the scope for constructing working MOND theories is very wide. Only a small corner of this scope has been explored so far.
\par
Some examples of these predictions are as follows:

a. Scale invariance in the DML implies that asymptotically far from any isolated mass, orbital speeds of test particles become independent of the orbital size, and depend only on the central mass, the shape of the orbit, and the relative location on the orbit. This gives, as a special case, asymptotically constant rotational speeds [asymptotically flat rotation curves of disc galaxies: $V\_{{\rm rot}}(r)\rar V\_{\infty}$]. That this prediction is confirmed by the data may be viewed as the observational epitome of SI.

b. The acceleration basis of MOND, with SI, predicts that $V\_{\infty}\propto (M\azg)^{1/4}=c(M/M\_M)^{1/4}$. It is the convention to normalize $\az$ so that equality holds.

c. A tight correlation is predicted between the locally measured acceleration, $g$, and mass discrepancy, $\eta$: $\eta\approx 1$ (no discrepancy) for $g\gg\az$, $\eta$ beginning to depart from 1 around $\az$, and $\eta\sim \az/g\gg 1$ for $g\ll\az$.

d. $\S\_M\equiv \az/2\pi G$, is a MOND critical surface density that is predicted to appear in various galaxy-dynamics correlations.

e. Scale invariance of the WFL also implies predictions analogous to a and b above for light bending: the bending angle, $\theta$, produced by an isolated mass $M$ becomes asymptotically independent of the impact parameter, and this constant angle, $\theta\_{\infty}$, can depend only on $c^{-2}(M\azg)^{1/2}$. If $\theta$ is first order in $c^{-2}$, then \beq \theta\_{\infty}\propto(M\azg/c^4)^{1/2}. \eeqno{lensa}
The proportionality factor depends on the theory.\footnote{It can even be $0$ if the MOND behavior is subdominant to the GR one, for example, in a theory where the MOND metric is conformally equivalent to the GR metric, as happens, for example, in the scalar-tensor toy relativistic MOND version discussed in Ref. \cite{bm84}.}

f. Even full rotation curves of spiral galaxies are largely predicted by the basic tenets alone.
\par
Some generalities regarding the above MOND predictions -- pertinent to theory construction -- are as follows:

1. Those predicted laws follow, by and large, from only the basic tenets of MOND \cite{milgrom14}, and so are  shared by any MOND theory that embodies these tenets.
Since these laws constitute the more straightforwardly testable predictions of MOND, it has been, so far, impossible to use them to distinguish between known MOND theories.

2. On the other hand, it is reassuring that there are so many strong and clear-cut predictions whose tests have helped establish the status of MOND as an alternative to DM, even without a final theory.

3. Several of these predictions involve $\az$ in different roles (in some appearing in the combination $\azg$, in some as $\az/G$, and in some alone). So, they can be used to determine $\az$ in several independent ways.
The fact that they all give consistent results is part of the general vindication process of MOND as a phenomenological scheme. But, more profoundly, this fact also points strongly to MOND being a genuine modification of dynamics, and point theories in this direction. This is similar to quantum dynamics providing the unifying framework for disparate phenomena -- such as the black-body spectrum, the photoelectric effect, the hydrogen-atom spectrum, superconductivity, quantum hall effect, etc. -- which, without such a framework would appear as unrelated phenomena that somehow involve the same constant $\hbar$.

4. These predictions are independent in the sense that they do not lead to each other in the framework of the DM paradigm.\footnote{One can construct families of baryon-plus-DM galaxy models that satisfy any subset of these predictions, but not the others.} This conclusion is further buttressed by noting that when interpreted within the DM paradigm, some of these predictions pertain to properties of the DM halo (e.g., a, d, and the first part of e, above), some to the baryons alone, and some to interrelations between them (e.g., b and the second part of e).
This too points strongly, to my mind, to MOND being underlain by a new theory of dynamics and not being merely a parsimonious summary of what DM does (as some DM advocates would have it).
\par
But of course, these predicted general laws, extensive as they are, do not replace a theory. We need a theory because there are many questions that these laws do not answer, and many, more detailed, predictions that do depend on the exact theory.

\section{\label{azero}Significance of the MOND acceleration constant}
It is now well established that an acceleration constant, $\az$, underlies, and appears ubiquitously in, disparate phenomena and regularities that characterize the observed dynamics of galaxies (e.g., Refs. \cite{milgrom83,sanders90,mcgaugh04,scarpa06,tiret09,
milgrom09b,fm12,milgrom14,trippe14,wl14}). Such a constant does not emerge in any known version of the DM paradigm.
\par
It is found that $\az\approx (1.2\pm 0.2)\times 10^{-8}\cmss$. It has been pointed out from the very advent of MOND \cite{milgrom83,milgrom89,milgrom94} that this is a cosmologically significant acceleration. We have the near equalities
\beq \baz\equiv 2\pi \az\approx cH_0\approx c^2(\Lambda/3)^{1/2}, \eeqno{coinc}
where $H_0$ is the Hubble constant, and $\Lambda$ the observed equivalent of a cosmological constant.
\par
Equivalently, for the MOND length we have $\lm\approx 7.5\times 10^{28}\cm$, which is of the order of today's Hubble distance:
$\lm\approx 2\pi \ell\_H$ ($\ell\_H\equiv c/H_0$), or of the de Sitter radius, $\ell\_S$, associated with $\Lambda$: $\lm\approx 2\pi \ell\_S$. The MOND mass, $M\_M\approx 10^{57}\gr$,
is then $M\_M\approx 2\pi c^3/GH_0\approx 2\pi c^2/G(\Lambda/3)^{1/2}$, of the order of the closure mass within today's horizon.
\par
Thus, to the known, but unexplained, ``coincidences'' that underlie the mass-discrepancies in the Universe\footnote{These are: the fact that the baryon density is of the same order as the required density of cosmological DM, even though the two components are believed to have been determined at very different times, and by different mechanisms; and, the fact that the density of ``dark energy'' and of matter in the Universe are today of the same order.} MOND phenomenology has exposed another one: the appearance of the cosmological acceleration parameters in local dynamics in systems much smaller than the Universe.
\par
This numerical coincidence, in itself, has some ``practical'' phenomenological ramifications (discussed, e.g., in Ref. \cite{milgrom14}), for example, that so-called ``strong gravitational lensing'' of cosmological sources cannot occur in the MOND regime.
It also has important implications for theory building. First, it points to the possibility that $\az$ varies with cosmic time, because the significant coincidence might be between $\az$ and the Hubble parameter, which varies with time.\footnote{Such variations could be detected by analyzing MOND dynamics (for example, the intercept of the mass-velocity relations) at different redshifts.}
\par
This coincidence also implies that we cannot have local systems that are both relativistic and in the MOND regime; for example, we cannot have a black hole that probes the MOND regime. For a system of mass $M$ and size $R$, being relativistic implies that $MG/R\sim c^2$, while being in the MOND regime implies $MG/R^2\lesssim\az$. Together, these imply
that $R\gtrsim\lm>\ell\_H$; so the system cannot be smaller than today's cosmological horizon.
\footnote{This holds in the past as well, even if $\az$ varies according to $\az\approx  cH/2\pi$ to preserve the first of the near equalities in eq. (\ref{coinc}).}
Thus, there are no systems, except the Universe at large, for which we encounter both DML and relativistically-strong conditions. This means, on one hand, that we cannot get observational clues or constraints on the DML, strong-field limit of a relativistic MOND theory, except from cosmology.
It also means that for the description of all phenomena, except cosmology, we need only a WFL theory.
\par
Beyond its ``practical'' implications,
this ``coincidence'' may be an important pointer in constructing MOND theories. If indeed fundamental, it may bespeak the most far-reaching implication of MOND: The state of the Universe at large strongly enters, and affects, local dynamics of small systems. And so, theories may be sought that implement this connection.\footnote{Mach's principle, whereby inertia is due to the influence of far away matter in the Universe, endorses such a concept.}
\par
Specifically, this cosmology-redolent value of $\az$ may explain why it is an acceleration that marks the boundary between two disparate dynamics \cite{milgrom99}: It is well known, and shows up in various contexts, that an acceleration $a$ defines a physically significant scale length $\ell_a=c^2/a$. For example, this is the radius of the near field of an accelerated charge; it is the distance to the event horizon carried by a uniformly accelerated system; it is the characteristic wavelength of the Unruh radiation associated with an accelerated system, etc. In a sense, then, low accelerations probe large distances.
Thus, for $a\gg\az$ we have $\ell_a\ll\lm$, so $\ell_a$ does not probe cosmologically significant distances, whereas for $a\ll\az$ it does.\footnote{In more detailed considerations of this kind we have to reckon with questions such as: ``Which of the cosmological acceleration parameters ($cH_0$, $c^2(\Lambda/3)^{1/2}$, or another) is $\az$ related to?'' and ``Does $\az$ vary with cosmic time, and if it does, how?''.}

\section{\label{dml}The deep-MOND limit}
The DML -- where MOND departs the most from standard dynamics -- is pivotal for understanding both MOND phenomenology and the requirements from MOND theories; so I discuss it now in some more detail.
\par
Start from a theory of gravitational dynamics -- involving as constants $G$, $\az$, possibly $c$, and masses -- which we want to examine as a candidate MOND theory. Its DML may be formally obtained by applying a space-time scaling to all the degrees of freedom (DoFs) in the equations of motion, $(t,\vr)\rar\l(t,\vr)$, and letting $\l\rar\infty$. (The DoFs may have nonzero scaling dimensions; for example, a scalar field may transform as $\psi(\vr,t)\rar\l^\xi\psi(\vr/\l,t/\l)$, where $\xi$ is the scaling dimension of $\psi$.)\footnote{ Note the appearance of $\l$ in the denominator for the independent variables -- a possible source of confusion. We can assign to DoFs other than length and time scaling dimensions at will; they need not match their units dimensions.}
If the limit exists as a consistent theory, it is automatically SI, as further scaling by a finite factor $\l$ has no effect.  In such a limit, all the DoFs with dimensions of acceleration scale as $g\rar\l^{-1}g\rar 0$, and so in the limit indeed  $g\ll\az$, as required in the DML (note that the constants of the theory, such as $G,~\az$ are not affected by the scaling). If the limit is also nontrivial -- in the sense that it retains the physics we want to describe -- the theory is a good candidate MOND theory. This is not the case for standard dynamics, for example, where the limit can only describe a theory with zero masses.
\par
In light of this it may well be that the transition between standard dynamics and MOND is achieved by a renormalization flow of some master theory; but this is yet to be demonstrated.
\par
An equivalent route to the DML is to scale only the constants of the theory according to their dimensions, as follows: If a constant $q$ has dimensions $[q]=[l]^a[t]^b[m]^c$, then $q\rar \l^{-(a+b)}q$; then let $\l\rar\infty$.\footnote{In this procedure we assume, without loss of generality, that the DoFs are normalized (e.g., by multiplying them by a power of $\az$) so that their scaling dimension matches their $[l][t][m]$ dimensions: i.e., if $[\psi]=[l]^\b[t]^\c[m]^\z$, then its scaling dimension is $\xi=\b+\c$. With this normalization, a scaling transformation is equivalent to a change in the constants alone, corresponding to a scaling of the length and time units by the same factor.} In our context we have $c\rar c$, $G\rar \l^{-1}G$, $\az\rar\l\az$, and $m_i\rar m_i$. So, $\az\rar\infty$, as befits the DML, $G\rar 0$, but $c$, $m_i$, and $\azg= G\az$ remain fixed, and they are the only constants that can remain in the DML. For more details see Ref. \cite{milgrom14}.
\par
We recapitulate that it is the presence of $\az$ in MOND that allows us to get an interesting SI limit for purely gravitational theories with finite masses. In a theory with only $G$ and masses (and possibly $c$) as constants, the fact that $G$ has dimensions $[G]=[\ell]^3[t]^{-2}[m]^{-1}$ stymies SI. We could make $G$ invariant if we accompany the space-time unit scaling by one of the mass units as well, but then masses are not invariant, unless they all vanish.\footnote{Alternatively, work in units where $G=1$, but then masses have units of $[\ell]^3[t]^{-2}$, and are not invariant.} With $\az$ in, we could form a meaningful SI theory with finite masses, by employing $\azg$.
\par
In strict DML dynamics there is, thus, neither $\az$ nor $G$, only $\azg$ appears. It is the existence of a Newtonian regime of phenomena that necessitates introducing $G$ and $\az$ separately. Below, I shall use $\azg$ in deep-MOND results.
\par
Because only $\azg$ appears, apart from masses, in the NR DML, we infer (e.g., \cite{milgrom09a}) that the dynamics it describes is invariant, more generally, to all scalings of the two-parameter family $l\rar\a l$, $t\rar\b^{-1} t$, $m\rar (\a\b)^4 m$ (because the analogous change of units leaves $\azg$ intact). For example, if a system of masses $m_i$ having orbits $\vr_i(t)$ (such as a galaxy) is a solution of the DML equations, then so is the system of masses $\hat m_i=(\a\b)^4 m_i$ having orbits $\hat\vr_i(t)=\a\vr_i(\b t)$, and thus velocities  $\hat\vv_i(t)= \a\b\vv_i(\b t)$ (with the appropriately scaled initial conditions). This leads to various MOND predictions whereby masses scale as the fourth power of velocities.\footnote{In a NR theory built, unlike MOND, for example, around a new length constant, $\ell\_0$, instead of an acceleration, SI in the limit of large distances, $\ell\_0\rar 0$, would imply that only the ratio $G/\ell\_0$ can appear in the limit. In this case the two-parameter family of invariances would have had $m\rar (\a\b)^2 m$.}
\par
It is interesting to imagine how a strict DML world would be like. Remaining still in the framework of pure gravity WFL, we see that in such a world, masses lose their role qua gravitational couplings (in this capacity they appear always as $MG$, and we are in the limit $G\rar 0$) \cite{milgrom09a}. But, of course, it is not gravity that disappears. The exact dynamics in a gravitational DML system depends on the particular MOND theory we employ. But we can point to some interesting general facts (results of the basic tenets) that illuminate the situation.
\par
Take any gravitation-producing body. We could define and measure for such a body an attribute with the dimensions of velocity, $V\_M$ -- our name for $(M\azg)^{1/4}$ -- which determines the asymptotic gravitational properties of the body (we can use $V\_M^4$ instead). It is the asymptotic rotational speed {\it of test particles} on circular orbits around the body when it is isolated from other bodies. It can also be defined from the constant, asymptotic  light-bending angle.\footnote{This is similar to the definition of the mass of a body in GR by the behavior of its asymptotic field.}
Note also, that $\azg$ itself does not appear in a DML world by  itself, nor do masses, only quantities that in our world we would identify as $M\azg$, or the densities of such quantities. Thus, $c$ is the only ``constant of nature'' we are left with, and a given system may further be characterized by $V\_M$ of various bodies in it.
\par
The facts that such a quantity can be defined -- i.e., that the asymptotic rotational speed is constant, and that there is a related, constant light-bending angle -- are nontrivial consequences of the new physics; they are the expression of the MOND laws (a) and (e) listed above in Sec. \ref{salient}. Furthermore, we may wonder then why for the Universe at large we measure $V\_M\sim c$, which is the coincidence expressed in eq.(\ref{coinc}).
In such a world we would also notice the significant fact that the $V\_M$ value of a composite body is related to those of its constituents, in that the $V\_M^4$ values are additive.\footnote{Barring the contribution of the interaction to the total mass.} This reflects part of the MOND law (b) above.
(Note that this does not refer to the full dynamics of a composite system, which cannot be described simply, and in a theory-independent way, in terms of the dynamics of its constituents. It is only a relation between some asymptotic attribute of the dynamics.)
\par
All of the above concerns the MOND laws that pertain to DML physics.
In our real world, however, MOND predicts a rather richer variety of laws. In our world, we can measure both the standard dynamics aspects of a body, such as a galaxy; e.g., its total (baryonic) mass, M,\footnote{This can be done either dynamically, if the body is a high-acceleration one (i.e., contained within its MOND radius), so it has a standard dynamics regime around it, or even if it is itself a DML object, but we can add up the masses of its constituents (e.g., stars), which are intrinsically of high acceleration.} and its DML attributes, e.g., $V\_M$,  and MOND predicts relations between them. For example, since these two attributes of a body are apparently independent, the fact that we find the ratio $V_M^4/M$ ($=\azg$) to be a universal constant [law (b)] is highly nontrivial. In other words, we notice that $V\_M$ of a body also determines its Newtonian dynamics.
\par
All these only express in a different way what was already said before, but it is useful to see things from this angle as well.

\section{\label{structure}MOND theories --generalities}
\par
It seems to me that in understanding MOND and its fundamentals we have only scratched the surface. If the developments of quantum mechanics and relativity are any lesson here, departures of such magnitude from long- and well-tested physics may bring with them completely new concepts, not summarized by mere modifications of the governing actions or the equations of motion.
MOND may also turn out to bring in concepts that are presently beyond our ken (as hinted perhaps by the cosmological connotations of $\az$).
\par
When coming to construct MOND theories, a natural question one asks is: ``which of the cherished principles that underlie standard dynamics --  such as Lorentz invariance and other symmetries, the various equivalence principles, general covariance, etc. -- should be retained in a MOND theory?''
\par
These principles of standard dynamics are largely based on experiment and observation.
But, observations on scales of galaxies and above completely disagree with standard dynamics anyhow (barring their rescue by dark matter), at least, this is the very starting point of the MOND paradigm. And, in any event, those underlying principles have not been tested in the region of low accelerations that is of relevance to MOND; so they may be strongly broken for $g\lesssim\az$, while well obeyed for $g\gg \az$.
\par
Moreover, there are strong signs that MOND, as we now perceive it, is an effective, approximate theory.
If so, clearly some principles, even if obeyed by the more fundamental theory, might be spontaneously  broken by the effective MOND theory that we may have to satisfy ourselves with temporarily.
\par
For example, the omnipresence of the thermal cosmic microwave background clearly breaks Lorentz invariance (although dynamically unimportant, and irrelevant for MOND). In an analogous vein, it has been suggested in connection with MOND \cite{milgrom99} that the quantum vacuum might determine inertia, and in a nearly de Sitter universe, such as ours, might lead to MOND dynamics. We know, at least, that the vacuum defines an inertial frame such that accelerated observers can measure their acceleration via an Unruh-like effect, but here too no dynamical effects leading to MOND have been demonstrated.
\par
This, and other radical concepts have come up in attempts to construct MOND theories.
Such approaches have their attractions, and, justifiably, continue to be explored,
but, as far as I can judge, they have not yet led to full-fledged MOND theories: theories that can be applied, in principle, to any physical system, and that obey certain consistency requirements, for example, the standard symmetries, causality, etc.
Such ideas for theories will be discussed in Sec. \ref{ideas}.
\par
The presently known, full-fledged MOND theories are qualitatively rather mundane, if quantitatively drastic modifications of GR or of its NR limit. These theories add new DoFs, and modify the underlying action; but, they do not deeply depart in spirit from their predecessors.\footnote{Such theories might be analogous to the pre-GR attempts to relativitize  gravity by writing Lorentz-invariant theories for the gravitational potential -- as, for example, in Nordstr\"{o}m's theories -- before the full force of gravity-as-space-time-geometry was appreciated.}
\par
Take GR as a benchmark and starting point.
Its action, governing a pure-gravity system of masses $m_i$, is
\beq I\_G=I_g+I_p,  \eeqno{vytres}
where the ``free'' action for the metric is the Einstein-Hilbert one (hereafter I adopt units where $c=1$)
\beq I_g=-\frac{1}{16\pi G}\int\gh Rd^4x, \eeqno{lulap}
and the particle action is
\beq  I_p= -\sum_i m_i\int d\tau_i. \eeqno{vyutda}
In the WFL of GR we write $\gmn=\eta\_{\m\n}+h\_{\m\n}$, where $\eta\_{\m\n}$ is the Minkowski metric, and the theory is treated to lowest order in $h\_{\m\n}$. So,
\beq  I_g\approx -\frac{1}{16\pi G}\int \bar E(h\_{\m\n}) d^4x,  \eeqno{nasarat}
where
the lowest order Einstein-Hilbert Lagrangian is
\beq \bar E(h\_{\m\n})=\frac{1}{4}[{{h\^{\n\r}}\_{,}}\^\c(h\_{\n\r,\c}-2h\_{\n\c,\r})
-{{h}\_{,}}\^\c(h\_{,\c}-2h\^{\r}\_{\c,\r})], \eeqno{nomas}
and,
\beq I_p\approx -\sum_i m_i\int[(\eta\_{\m\n}{\dot x}\_i\^\m {\dot x}\_i\^\n)^{1/2}+\frac{1}{2}(\eta\_{\s\r}{\dot x}\_i\^\s {\dot x}\_i\^\r)^{-1/2}h\_{\m\n}{\dot x}\_i\^\m {\dot x}\_i\^\n] d\tau_i,
  \eeqno{gepada}
where ${\dot x}\_i\^\m=dx\_i\^\m/d\tau\_i$.
\par
In the fully NR limit, where we consider only slow moving particles, we can write
\beq I=\int L~dt,~~~~~~~ L=\int \L~\drt.  \eeqno{miot}
 \beq \L_g=-\frac{1}{8\pi G}(\gf)^2,~~~~\L_p=\frac{1}{2}\r \vv^2 -\r\f, \eeqno{manare}
where the continuum version is now used, with $\r(\vr,t)=\sum\_im\_i\d[\vr-\vr_i(t)]$, and similarly for the velocity field. Particle motions are determined from $\va=-\gf$, while the gravitational potential, $\f$, which stands for $-h\_{00}/2$, is determined from the Poisson equation sourced by $\r$.
\par
In the minimalistic approach, one then modifies one of these three versions of standard dynamics by
introducing $a_0$, possibly adding new DoFs, and constructing a MOND action that satisfies the basic premises of MOND. Existing theories involve an interpolating function that is put in by hand to artificially interpolate between the MOND and the high-acceleration regimes:
The theory defines a certain scalar functional, $A$, of the DoFs, having the dimensions of acceleration, and introduces in the action a function $\F(A/\az)$, such that for $\az\rar 0$ the theory tends to the standard one, while in the opposite limit it gives an adequate DML, namely one whose equations of motion are SI.
\par
The appearance of an interpolating function in the action of these theories  --  together with the above-mentioned coincidence of $\bar a_0\approx c^2(\Lambda/3)^{1/2}$ -- is to my mind, a strong indication that they are all, at best, approximate, effective theories of limited applicability.
\par
``Interpolating functions'' that connect the classical limit with the relativistic or the quantum limit, occur also in the context of relativity or quantum physics.
In relativity: the Lorentz factor, the particle dispersion relation $E=(p^2+m^2)^{1/2}$ ``interpolating'' between $E\approx p$ and $E\approx m+p^2/2m$, various relations describing dynamics around black holes, which interpolate between the near- and  asymptotic-region behaviors, etc. In the classical-quantum connection:  the Planck black-body function interpolating between the classical, Rayleigh-Jeans, and the high-frequency behaviors, the specific heat of solids, interpolating between the high-temperature, classical, Dulong-Petit expression and the quantum-mechanical behavior at low temperatures, etc. However, these ``interpolating functions'' each pertain to a specific phenomenon, and may differ in meaning and form for different phenomena. Also, they are derivable from the theory, not appearing in its action.
\par
We expect such diverse interpolating functions to appear in a more fundamental MOND framework, and here too they would be specific to the phenomenon at hand: perhaps one for circular orbits, another for linear, constant-acceleration trajectories, yet another related to the external field effect, etc. But they would all, of course, correspond to SI in the DML. This would clearly happen, for example, in a MOND theory based on vacuum effects, alluded to above, where the Unruh radiation strongly depends on trajectories.
\par
These existing MOND theories are nonetheless very useful, having provided proofs of various concepts: First, that MOND theories can be written that are derived from an action, satisfy all the standard conservation laws, give the correct center-of-mass motion of a composite body, etc. Then, that covariant theories can be written with, for example, correct lensing. They also provide consistent frameworks for conducting various calculations, being complete theories that make the major predictions of MOND correctly \cite{milgrom14}. Also, at least on galactic scales, they make very similar, if not always quite the same, predictions of more detailed aspects such as full rotation curves of spiral galaxies \cite{brada95,angus12,milgrom12c}, or of gravitational lensing.
Still, we do not know how far to trust them beyond this.
\par
There are many ways to achieve SI of the DML equations of motion (see below).
In fact, the mere realization that this ensures the salient MOND predictions, shows how large the scope of potential theories is.
\par
Scale invariance of the equations of motion does not require that the action be SI; it is enough that it have a well-defined scaling dimension, $\xi$, namely that under scaling $x^\m\rar\l x^\m$, we have $I\rar\l^\xi I$. This is not the case for standard dynamics, as $I_g$ has scaling dimension $\xi=2$, while $I_p$ has $\xi=1$.
\par
We can construct a ``MOND theory'' (one that makes the basic MOND predictions) by modifying one or more terms in the standard action to obtain one with a well-defined scaling dimension. This may serve as a trial DML for a MOND theory. It then has to be ascertained that a theory exists that has this action as its limit for $\az\rar\infty$, and that restores standard dynamics in the opposite limit.
\par
We could modify only $I_g$ (modified gravity) to make its DML have $\xi=1$, modify only $I_p$ to make its DML have $\xi=2$, or, we could modify both to have the same DML scaling dimension, perhaps even to have $\xi=0$, so the DML of $I\_G$ is SI. There are more possibilities, some of which are exemplified below.
\par
The role of SI had not been recognized at the time existing MOND theories were advanced.
Instead, what was imposed is essentially asymptotic flatness of rotation curves. But now we have a rather wider view of this aspect of MOND, and we see that
these theories can all be described in terms of SI (see more below).

\subsection{\label{modgrav}Modified-gravity MOND}
In one large class of modifications, called modified-gravity (MG) MOND theories, the theory can be formulated such that the matter action, $I_p$ remains intact and only $I_g$ is modified. In the relativistic case, the Einstein-Hilbert action is modified to a MOND action, possibly involving additional DoFs to the metric, but of all these DoFs, matter couples directly only to the metric.\footnote{There might then be equivalent versions of the theory where the matter action is modified. For example a scalar-tensor theory, such as the Brans-Dicke theory, may also be formulated with a metric that satisfies the Einstein equation, while matter is minimally coupled to another (``physical'') metric, conformally related to it.}
Most of the existing relativistic MOND theories involve a Riemannian metric, but, for example, the theory proposed in Refs. \cite{chang08,lichang12} is based on Finslerian geometry.
\par
Because $I_p$ is left intact, and it has scaling dimension 1, the DML of the modified $I_g$ must also have this scaling dimension. This is indeed seen to be the case with the MG, MOND theories discussed below.
\par
In local theories -- unlike, e.g., the nonlocal MOND theory in Ref. \cite{deffayet11} -- it is not possible to construct an acceleration tensor, $A$, from the metric alone (since such quantities must vanish in locally flat coordinates, and hence identically). Thus local, relativistic MOND theories involve, perforce, additional gravitational DoFs besides the metric. In WFL or NR theories, where the background metric is a priori fixed to be Minkowskian (so we lose much of the coordinate freedom), we can construct acceleration quantities (from $h\_{\m\n}$ or the potential $\f$); so at this level there is no need for extra gravitational DoFs.
\par
In the NR limit of such theories, only the Poisson Lagrangian [$\L_g$ (\ref{manare})] is modified, possibly to include additional DoFs, but the gravitational potential $\f=-h\_{00}/2$ is singled out of all the gravitational DoFs, in that it alone determines the motion of matter via $\va=-\gf$.
\par
While a NR problem may be time dependent, the time evolution is conveniently decoupled from the solution of the gravitational-field equations. These do not involve time derivatives and so are solved as a static problem for each time separately, given the source $\r$ at each time.
\par
In the DML of such a theory, SI of the particle equation of motion, $\va=-\gf$, implies that the scaling dimension of $\f$ has to be $0$; namely, under scaling, $(\vr,t)\rar\l(\vr,t)$, we have $\f(\vr,t)\rar\f(\vr/\l,t/\l)$.
Thus, SI of the gravitational-field equations (which involve no time derivatives) implies that they are invariant to space dilatations, under which $\vr\rar\l\vr$, $\r(\vr)\rar\l^{-3}\r(\vr/\l)$.
\par
It was shown in Ref. \cite{milgrom14a} that this leads to a very useful virial relation that applies to an isolated, self gravitating, DML system of point-like masses, $m_i$, at positions $\vr_i$, subject to MOND forces $\vF_i$. This relation reads
 \beq \sum_i \vr_i\cdot\vF_i=-\frac{2}{3}\azg^{1/2}[(\sum_i m_i)^{3/2}-\sum_i m_i^{3/2}]. \eeqno{i}
This relation then holds in all MG, MOND theories (subject to some very plausible assumptions).
Important corollaries of this relation are an analytic expression for the gravitational force between any two masses, in the DML, and a virial relation that determines the velocity dispersion of a DML system of point-like bodies in terms of their masses alone. \cite{milgrom97,milgrom14a}

\subsection{Modified-inertia MOND}
Modified inertia is a generic name given to MOND theories that are not of the MG type. This is because here we modify the particle free action (and possibly other matter actions -- see Sec. \ref{dmlem}), which encapsulates the inertial properties of matter.
\par
This approach may require some drastic changes in the way we describe the behavior of matter. For example, it may require abandoning Lorentz invariance, perhaps for another symmetry (see Sec. \ref{ideas}). Or, it may entail a description of matter with different, or additional, DoFs to those we now employ.
\par
If we hark back to historical examples, special relativity may be viewed as modified inertia compared with Newtonian dynamics; indeed, it entailed a change of the basic dynamical symmetry. Quantum mechanics, among other changes, modified our very description of matter, and introduced spin as an additional DoF. String theory also changes the way we describe the elementary constituents of matter, from point-like objects to extended ones, requiring a different set of DoFs to describe matter.
\par
Another example, particularly interesting in the present context:
It has been speculated in recent years that a yet-hidden sector exists, of entities, dubbed unparticles \cite{georgi07},
whose kinematics are SI in the quantum context -- namely, they bypass the $\hbar$ obstacle to SI  --  without banishing the effects of finite mass (their mass is rather undefined by the standard view of mass, in that they do not have  a fixed $E^2-p^2$).\footnote{In the quantum context, energy and momentum have to have scaling dimension $-1$ (as $E=\hbar\n$, etc.); so $E^2-p^2=m^2$ is not SI.}
\par
Because $I_g$ might be considered the free action of the metric, it is conceivable that it too has to be modified in modified-inertia theories.
\par
Our aim is, as we saw, to construct a modified DML in which all terms in the action have the same scaling dimension. This could be done, for example, by modifying $I_p$ to have $\xi=2$ like $I_g$, or modifying both to have the same $\xi$.
\par
Even more possibilities open if we note that we also have some freedom to change at will the scaling dimensions of some of the DoF.
We saw that looking for a MOND theory, we may be justified in starting from the WFL of GR, eqs.(\ref{nasarat})(\ref{gepada}), or from the NR limit, eqs.(\ref{miot})(\ref{manare}).  We can then modify only one of the terms in $I_p$ and change the dimension of others without actually modifying them.
\par
For example, in the NR formulation, eq.(\ref{manare}), give $\f$ scaling dimension of -1.  Then, the action $\int (\L_g-\r\f)d^3rdt$ has $\xi=0$, namely it is SI.
The same is true, more generally, of the WFL of $I_g$ in eq.(\ref{nasarat}), and the second term in expression (\ref{gepada}), if we give $h\_{\m\n}$ dimension $-1$.
But $\int\r\vv^2 d^3rdt$ [or the first term in expression (\ref{gepada})] has $\xi=1$, and should be modified to also have a SI, DML.
\par
This particular route has been discussed in detail in Ref. \cite{milgrom94} (not in terms of SI).
We see that the gravitational potential is still determined from the
Poisson equation. If we define $\psi =\az\f$, we have $\Delta\psi=4\pi\azg\r$, which is SI, $\psi$ having dimension $\xi=-1$.
The particle equation of motion is now of the form
 \beq \textbf{A}[\{\vr(t)\},\az]=-\az^{-1}\grad\psi,
  \eeqno{huta}
instead of $\ddot\vr=-\az^{-1}\grad\psi$. $\textbf{A}$ is a functional of
the whole trajectory $\{\vr(t)\}$, with the dimensions of acceleration. For $\az\rar 0$, $\textbf{A}\rar \ddot\vr$. In the opposite limit $\textbf{A}$ attains scaling dimension $-2$.
\par
It was found in Ref. \cite{milgrom94} that
if such an equation of motion is to follow from an
action principle, enjoy Galilei invariance, and have the above Newtonian and MOND limits, it has to be time nonlocal. But it may be that with a more general choice of DoFs to describe particles, or with a change of symmetry, local theories can be found.
\par
We do not yet have a fully acceptable theory in this vein, even in the NR regime. Only some toy
theories have been partly explored
\cite{milgrom94,milgrom11}.
\par
An important and robust prediction shared by all theories
in the class is: For circular trajectories in an axisymmetric
potential, eq.(\ref{huta}) has to take the form
 \beq \m\left(\frac{V^2}{r\az}\right)\frac{V^2}{r}=
 -\frac{d\f}{dr}.  \eeqno{rc}
Here, $V$ and $r$ are the orbital speed and radius, respectively; and $\m(x)$ is
universal for the theory, but applies only to the description of circular trajectories, and is derived from the expression of the
action specialized to such trajectories. We have $\m(x\ll 1)\approx x$,
$\m(x\gg 1)\approx 1$. It is this relation that has been used in most MOND
rotation-curve analyses to date.
\par
It may seem unjustified to modify only part of what in standard relativistic physics arise from a single particle action. However, exactly this sort of modification is expected, for example, in effective theories in which inertia is modified due to some interaction with an agent that does not modify the gravitational properties.
For example, I already mentioned the heuristic idea  \cite{milgrom99} that MOND, indeed
inertia itself, can result from an effect of the vacuum, (where the origin of $\az$ in cosmology also emerges). The vacuum then serves as an absolute inertial
frame (acceleration with respect to the vacuum is detectable, e.g., through the Unruh effect). Such effects may well modify only the kinetic part of the particle action without modifying the gravitational part.
\par
There are, in fact, many known examples of this happening in effective theories. For example, electrons in solids may behave effectively as free particles with modified inertial (but not, of course, gravitational) properties, where the inertial mass changes and can even become anisotropic. So the $\r\vv^2/2$ Lagrangian is modified, but not the $\r\f$ one.
It is also known that the interaction of the electromagnetic field with
charged vacuum fields modifies the free action of the
electromagnetic field, by the so-called Heisenberg-Euler effective action (see e.g.  \cite{izu} p. 195).
\par
It is not clear whether, and exactly how,
specific such mechanisms can produce MOND effects.
Some suggestions to this effect are mentioned in Sec. \ref{ideas}
\section{\label{theories}Existing MOND theories}
In this section I list and discuss briefly the main, full-fledged MOND theories constructed to date. They all follow the general schematics detailed in Sec. \ref{structure} above.
\subsection{Nonrelativistic theories}
\subsubsection{Modified Poisson gravity} In this theory \cite{bm84}, the Newtonian gravitational action [$\propto(\gf)^2$] in eq.(\ref{manare}) is modified into
\beq \L_g=-\frac{\az^2}{8\pi G}\F[(\gf)^2/\az^2].\eeqno{jufra}
 Thus, the
Poisson equation for the gravitational potential is replaced by a nonlinear version
\beq\vec \nabla\cdot[\mu(|\vec\nabla\phi|/\az)\vec\nabla\phi]= 4\pi G \rho, \eeqno{poissona}
with $\m(x)\equiv \F'(x^2)$.
We saw that the DML field equation for $\f$ has to be dilatation invariant, which implies $\m(x\ll 1)\propto x$ (equality is imposed by the normalization of $\az$).
\par
in a pure-DML language, a continuum gravitating system is described  by the distribution of velocities $\vv(\vr,t)$ and of the quantity $\eta(\vr,t)$ that represents $\azg\r(\vr,t)$, whose dimensions are of density of velocity to the fourth power. The DML Lagrangian of such a system can be taken as
 \beq -\frac{1}{12\pi}[(\gf)^2]^{3/2}+
 \eta(\frac{1}{2}\vv^2-\f)  \eeqno{dasa}
 (its scaling dimension is $\xi=-3$, $\f$ having $\xi=0$),
and the field equations are
 \beq \dot\vv=-\gf,~~~~~~
 \vec\nabla\cdot(|\vec\nabla\phi|\vec\nabla\phi)= 4\pi \eta. \eeqno{ddmm}
\par
Very interestingly, this equation is not only dilatation invariant, but, more generally,
invariant under space conformal transformations \cite{milgrom97}: Namely, beside its obvious invariance to translations, rotations, and dilatations, eq. (\ref{ddmm}) is invariant to inversion about a sphere of any radius $a$, centered at
any point $\vr_0$:
\beq \vr\rar\vR=\vr_0+\frac{a^2}{|\vr-\vr_0|^2}(\vr-\vr_0),
\eeqno{furat}
with $\f(\vr)\rar\hat\f(\vR)=\f[\vr(\vR)]$, and $\r(\vr)\rar\hat\r(\vR)=J^{-1}\r[\vr(\vR)]$, where $J$ is the Jacobian of the transformation (\ref{furat}).
This ten-parameter symmetry group of eq.(\ref{ddmm}) is thus the conformal group in 3-dimensional Euclidean space. It is the same as (isomorphic to) the isometry (geometrical symmetry) group of a 4-dimensional de Sitter space-time, with possible deep implications, perhaps pointing to another connection of MOND with cosmology \cite{milgrom09a}.

\subsubsection{Quasilinear MOND}
\par
Another theory, Quaslinear MOND (QUMOND) \cite{milgrom10a}, involves two potentials: $\f$, which alone governs the motion of masses through $\va=-\gf$, and an auxiliary potential $\fN$, which will turn out to equal the Newtonian potential for solutions of the field equations. In QUMOND, one modifies the gravitational Lagrangian in eq.(\ref{manare}) into
\beq \L_g=-\frac{1}{8\pi G}\{2\gf\cdot\gfN
-\az^2\Q[(\grad\fN)^2/\az^2]\}.\eeqno{juyta}
Thus, the
Poisson equation for the gravitational potential is replaced by the pair of field equations
\begin{equation}\Delta\fN= 4\pi G \rho,~~~~~~~~~\Delta\phi=\vec \nabla\cdot[\nu(|\vec\nabla\fN|/\az)\vec\nabla\fN],\label{eq:ii} \end{equation}
where $\n(y)=\Q'(y^2)$.
These require solving only the linear Poisson equation twice (hence the epithet quasilinear). In the DML, SI requires that $Q(z)\propto z^{3/4}$ ($\L_g$ then has scaling dimension $\xi=-3$, $\f$ having $\xi=0$ and $\fN$ having $\xi=-1$). Using the standard normalization of $\az$, and working with $\psi\equiv \az\phi_N$ (which makes the dimensions of $\psi$ match its scaling dimension), we have in the DML: \beq \Delta\psi=4\pi\azg\rho,~~~~~~\Delta\phi=
\vec\nabla(|\nabla\psi|^{-1/2}\nabla\psi). \eeqno{mier}
These are space-dilatation invariant -- as generally holds for such MG theories -- but, apparently, not conformally invariant.

\subsubsection{Generalizations}
The above two theories are special cases in a class of two-potential, MG theories \cite{milgrom10a}. These have a gravitational Lagrangian containing only first derivatives of the potentials, so it must be a function of the three scalars formed from $\gf$ and $\grad\psi$ ($\f$, as before, dictating motions)
  \beq \L_g=\L_g[(\gf)^2,
  (\grad\psi)^2,\gf\cdot\grad\psi]. \eeqno{manada}
\par
To obey SI, the DML of $\L_g$ has to be of the form
\beq \L_g\propto\sum_{a,b}s_{ab} [(\gf)^2]^\eta[(\grad\psi)^2]^\xi(\gf\cdot\grad\psi)^\theta, \eeqno{iiaa}
with the exponents related through $\eta=a+3/2$, $\xi=a+b(2-p)/2$, $\theta= b(p-1)-2a$; $p$ is fixed for a given theory and $a,~b$ are arbitrary.
The dimensions of $\f$ and $\psi$ are, respectively, $[l]^{2}[t]^{-2}$ and, if $b\not= 0$, $[l]^{2-p}[t]^{2(p-1)}$ (for $b=0$, the dimensions of $\psi$ are arbitrary); $s_{ab}$ are dimensionless. For any $p$, this reduces to the DML of the
nonlinear Poisson theory, $[(\gf)^2]^{3/2}$, when $a=b=0$. QUMOND is gotten for $p=-1$ with two terms with $a=-3/2,~b=1$  and $a=-b=-3/2$.  \footnote{Many of these theories may be unfit for various reasons.}
\par
For $p=0$ and any combination of $a,~b$, the DML of the gravitational field equations is conformally invariant. Likewise for $b=0$, in which case $p$ does not enter.

\subsection{Relativistic theories}
I now list briefly the main relativistic formulations known to date; some of these are discussed in detail in this volume. They are all based on a Riemannian metric as the main carrier of gravity as sensed by matter. Other options have been proposed, such as one based on a Finslerian geometry \cite{chang08,lichang12}.
\subsubsection{TeVeS}
The Tensor-Vector-Scalar (TeVeS), relativistic formulation of MOND,
has been put forth by Bekenstein \cite{bek04}, building on ideas by
Sanders \cite{sanders97} (see also reviews in Refs. \cite{skordis09,fs09}). TeVeS was the first full-fledged relativistic formulation of MOND. Its advent greatly helped advance the case for MOND, demonstrating for the first time that a decent covariant formulation is feasible.
 \par
In TeVeS, gravity is carried by a vector field,  ${\cal U}_\a$, and a scalar field, $\f$ (with their own free actions), in addition to a metric $\gab$ whose action is the standard Einstein-Hilbert action. However, matter
couple minimally not to $\gab$ itself, but to the ``physical'' metric
\beq \tilde g\_{\a\b}
=e^{-2\f}(\gab + {\cal U}_\a {\cal U}_\b) - e^{2\f} {\cal U}_\a
{\cal U}_\b. \eeqno{physic}
The auxiliary gravitational fields also couple to $\gab$. A Lagrange-multiplier term in the action constrains ${\cal U}_\a$ to be of unit length.
\par
TeVeS reproduces MOND phenomenology for galactic systems in the NR
limit, with a certain combination of its constants playing the role
of $\az$. In particular, when $\az\rar 0$, the NR limit goes to
Newtonian gravity. However, the relativistic theory itself does not exactly satisfy the second MOND tenet in that it does not go
exactly to GR at high accelerations. This remaining high-acceleration departure from GR has subjected TeVeS to constraints from the solar system ( e.g., \cite{sagi9}), and from binary compact stars \cite{freire12}. These constraints do not pertain to MOND aspects of TeVeS.
\par
As in GR, the potential that appears in the expression for lensing
by NR masses (such as galactic systems) is the same as that which
governs the motion of massive particles.
\par
Cosmology, the cosmic microwave background, and structure formation in TeVeS have been
considered in
\cite{dodelson06,skordisetal06,skordis06,skordis08,zlosnik08}. It
was shown that there are elements in TeVeS that could mimic
cosmological DM, although no fully satisfactory application of TeVeS
to cosmology has been demonstrated.
\par
Gravitational waves in TeVeS have been considered in
\cite{bek04,sagi10}.
\par
Galileon k-mouflage MOND adaptations \cite{babichev11} are said to help TeVeS avoid high-acceleration constraints.
\subsubsection{MOND adaptations of Einstein-Aether theories}
Einstein-Aether theories (e.g., \cite{jm01}) have been adapted to account for MOND phenomenology \cite{zlosnik07}. Gravity is
carried by a metric, $g_{\m\n}$, as well as a vector field, $A^\a$.
To the standard Einstein-Hilbert Lagrangian for the metric one adds
the terms \beq \L(A,g)=\frac{\az^2}{16\pi G}\F(\K)+
\L_L,\eeqno{lagrangeII} where
 \beq
\K=\az^{-2}\K^{\a\b}_{\c\s}A^{\c}\cd{\a}A^{\s}\cd{\b}.\eeqno{KKK}
$$\K^{\a\b}_{\c\s}=c_1g^{\a\b}g_{\c\s}+c_2\d^\a_\c\d^\b_\s
+c_3\d^\a_\s\d^\b_\c+c_4A^\a A^\b g_{\c\s},$$ and $\L_L$ is a Lagrange multiplier term that forces the vector to be of unit length. Matter is standardly coupled to $g_{\m\n}$.

The asymptotic behaviors of $\F$, at small and large
arguments, give the deep-MOND behavior and GR, respectively.
\subsubsection{Bimetric MOND gravity}
Bimetric MOND gravity (BIMOND)
\cite{milgrom09,milgrom10b,milgrom10c,cz10,milgrom14b} is a class of relativistic
theories, whereby gravity is described by two metrics, $\gmn$ and $\hgmn$. The Einstein-Hilbert action is replaced by the action
\beq I=-\frac{1}{16\pi G}\int[\b\gh R
+ \a\hgh \hat R
 -2(g\hat g)^{1/4}\az\^2\M] d^4x.
\eeqno{mushpa}
Here,  $R$ and $\hat R$ are the Ricci scalars of the two metrics ($c=1$ is taken). The metrics interact via the dimensionless scalar $\M$, which is a function of the two metrics and their first derivatives.
The novelty in BIMOND over earlier bimetric theories is in the choice of the interaction
term in accordance with MOND. The arguments of $\M$ are scalars built from the difference of the two Levi-Civita connections
 \beq C\ten{\a}{\b\c}=\C\ten{\a}{\b\c}-\hC\ten{\a}{\b\c},
  \eeqno{veyo}
which is a tensor that acts like the relative gravitational accelerations
of the two sectors. This is particularly germane in the context of
MOND, where, with  $\az$ at our disposal, we can construct from
$\az\^{-1}C\ten{\a}{\b\c}$ dimensionless scalars. The scalars constructed from the quadratic tensor
 \beq \Up_{\m\n}\equiv  C\ten{\c}{\m\l}C\ten{\l}{\n\c}
-C\ten{\c}{\m\n}C\ten{\l}{\l\c},  \eeqno{mamash} such as $\Up=\Gmn\Up_{\m\n},~~~\hat\Up= \hGmn\Up_{\m\n}$, have particular
appeal. There is also the standard matter action, and one for a putative twin matter, whose existence is
suggested (but not required) by the double-metric nature of the
theory. Matter DoFs couple only to the
standard metric $\gmn$, while twin matter couples only to $\hgmn$.
\par
BIMOND cosmology is preliminarily discussed in Refs.
\cite{milgrom09,cz10,milgrom10c}.
Some aspects of structure formation
are discussed in Ref. \cite{milgrom10c}.
The weak-field limit of BIMOND, in particular some aspects of gravitational-waves in BIMOND, has been treated recently in Ref. \cite{milgrom14b}.
BIMOND has several attractive features (shared by the modified Einstein-Aether theories):
It tends to GR for $a_0\rightarrow 0$; it has a simple NR limit; it describes gravitational lensing correctly; and, it has a generic appearance of a cosmological-constant term that is of order $a_0^2/c^4$, as observed. The DML of its weak-field limit is scale invariant \cite{milgrom14b}.
\subsubsection{Nonlocal metric theories}
Nonlocal metric MOND theories \cite{soussa03,deffayet11} are
pure metric, but highly nonlocal in that they involve operators that are functions of the 4-Laplacian. They agree with general relativity in the weak-field regime appropriate to the solar system, but possess an ultra-weak-field regime when the gravitational acceleration becomes comparable to $\az$. In this regime, the models reproduce the MOND force without DM and also give enough gravitational lensing to be consistent with existing data. It was proposed that these theories might emerge from quantum corrections to the effective field equations. A detailed account of this approach appears in this volume.

\subsubsection{Dipolar dark matter}
It was noted in Ref. \cite{blanchet07} that the analogy of eq.(\ref{poissona}) with the equation for the electric potential in a nonlinear dielectric medium may point to the possibility of a gravitationally polarizable medium giving rise to MOND phenomenology. This idea has been given a relativistic formulation \cite{blt08,blt09} that introduces a novel type of matter, dubbed ``dipolar DM'', which is gravitationally polarized by baryonic matter. The polarization then enhances the effective gravitational attraction of baryonic masses. By choosing an appropriate field potential -- equivalent to an interpolating function, and involving a constant that plays the role of $\az$ -- we can get eq.(\ref{poissona}) in the NR limit.
The constant $\az$, if it is the only new one allowed, also appears in a cosmological constant term, which might account for the MOND-cosmology coincidence.
\par
Another parameter of the theory controls the role of this novel medium as DM that acts gravitationally beside its polarization effect. It can, thus, double as cosmological DM \cite{blt09}. A detailed account of this theory is described elsewhere in this volume.

\section{\label{ideas}Additional theoretical schemes}
Many ideas have been suggested that depart from the above scheme for constructing MOND theories. Some do try to obtain an interpolating function from microscopic physics, some still put it in by hand. Some are more advanced, some less, but I think that none have yet led to a full-fledged theory.
Here I list briefly such ideas that show promise in some degree of obtaining MOND phenomenology. These can be classified into the following categories.
\par
There are ideas based on the description of gravity and/or inertia as emergent concepts of entropic and holographic arguments \cite{pikhitsa10,kt10,lc11,klinkhamer12,pa12,pazy13}.
This can be done on a de Sitter background, instead of a Minkowski one, and so, in a way that harks back to the discussion in Ref. \cite{milgrom99}, a MOND constant can appear naturally as related to $\ell\_S^{-1}$.
\par
Reference \cite{milgrom02} describes an approach in which our universe is a membrane embedded in a higher-dimensional space-time. In the NR formulation, the gravitational potential is an additional coordinate in the embedding ST, and $\az$ is a constant, external acceleration to which the membrane is subject. A theory governed by a Lagrangian as in eq.(\ref{jufra}) can be obtained by ascribing to the membrane an energy that depends on its geometry.
\par
Other ideas invoke DM with properly tailored properties (such as unusual interactions with baryons) so as to obtain MOND phenomenology in galaxies, and DM phenomenology in cosmology \cite{bruneton09,zl10,ho10,bettoni11,ho12}.
\par
Yet other ideas construct MOND adaptations of
Ho\v{r}ava gravity \cite{rom10,blanmars11,sanders11}, and other constructs \cite{trippe13,bernal11}.

\subsection{de Sitter symmetry}
Our physics hinges on local Lorentz invariance in 4-dimensional spacetime of Minkowskian signature. Thus, when including translation invariance, the ``fundamental'' spacetime symmetry is the 10-parameter Poincar\'e group, which reduces to the NR, Galilei-plus-translations group in the limit $c\rar\infty$. This has also been the starting point in all the attempts to construct MOND theories. It has been suggested \cite{bl68}, mainly on aesthetic grounds (see also Ref. \cite{dyson72}), that a more suitable fundamental symmetry might be the 10-parameter isometry group of a (4-dimensional) de Sitter spacetime. This idea has been much discussed in various contexts (see e.g., more recently Ref. \cite{aldrovandi09}). Unlike the Poincar\'e group, which discriminates between the 6 Lorentz rotations and the 4 translations, the de Sitter symmetry transformations are all treated equally (being all rotations in the flat 5-dimensional Minkowskian spacetime in which a de Sitter spacetime can be embedded). This is in the basis of the mathematical appeal of the de Sitter group, which, in addition, is more general. We may view it as parametrized by a length, $\ell\_S$, the de Sitter length, in addition to $c$. The algebra of the Poincar\'e group is then obtained as a reduction of the de Sitter algebra (corresponding to taking $\ell\_S\rar\infty$). Further reduction (affected by taking $c\rar\infty$) gives the standard NR group. But there are other reduction pathways to other algebras, classified in Ref. \cite{bl68}.
\par
There are reasons to suspect that this basic symmetry, and its various reduction paths, is of relevance in connection with MOND. The relevant hints, mentioned already above, are the fact that $\lm$ is of the order of the value of $\ell\_S$ that corresponds to the observed cosmological constant, and the fact that the de Sitter symmetry appears in the DML of some MOND theories.
\par
Despite much effort I have not yet been able to establish a firmer connection.
\section{\label{cosmology}MOND theory and cosmology}
With the advent of GR -- a theory designed to account for local phenomena -- is has also become possible to treat the whole observable universe as a particular system amenable to description within the theory.
This luxury can by no means be taken for granted in theories, good as they may be at describing local phenomena. It had not been the case with Newtonian dynamics, and it is largely not the case with quantum theory. Such inadequacy to describe the Universe at large may be due to a limited scope of the theory in question, to issues of boundary and initial conditions, which local theories require, etc.
\par
In the case of MOND, with its basic tenets, as described above, there is an additional
reason to doubt that they, and local theories built on them, can apply directly to cosmology. ``Coincidence'' (\ref{coinc}) points to MOND being somehow an effective, approximate theory, in which perhaps $\az$ does not even have a ``fundamental'' significance, but is a derived, effective constant, given that in the galactic context  we are dealing with systems much smaller than the Universe, and dynamical times much smaller than cosmological time.
\par
Examples of such effective theories that represent approximations to more fundamental ones under restricted circumstances, are rife in physics. Here is one that bears close analogy to what MOND might be like: Observations of physics very near the surface of the Earth instate Earth's free-fall acceleration, $g\_E$, as a ``fundamental'' constant. This ``constant'' is then noticed to be related to the seemingly unrelated radius of the Earth, $R\_E$, and the surface-hugging orbital speed around it $c\_E$: $g\_E=c\_E^2/R\_E$. We know that this is not a coincidence, but follows from better understanding of gravity, in general. And clearly, a constant, radial $g\_E$, while it describes near-surface phenomena well, cannot be used to describe Earth-gravity, in general.
In the context of MOND, we see $\az$ playing it's ``fundamental'' role in galaxy dynamics. We also note that it is related to the ``radius of the Universe", $R\_U$, and the fundamental speed, $c$, by $\az\sim c^2/R\_U$. This probably points to the fact that presently known MOND theories, which are based on the basic tenets, are not fit to describe cosmology. I already mentioned the additional indication for this, based on the fact that these theories involve an interpolating function that has to be put in by hand.
\par
The fundamental theory that would lead to MOND cosmology, and that gives rise to MOND-in-galaxies will have to be understood simultaneously with cosmology.
\par
There are several disparate ways in which such a connection between local dynamics and the state of the Universe at large can arise.
\par
For example, cosmology might be somehow determined by a yet unknown theory (which dictates the initial conditions, the appearance of ``dark energy'', of baryons, etc.), and then, due to connections we have so far missed in physics, the dynamics of small systems (e.g., galaxies) is affected by the state of the Universe in a way that gives rise to MOND, as an effective theory, with imprints (symmetries, the value of $\az$) from cosmology.
This is what happens, e.g, in the schemes studied in Refs. \cite{milgrom99,pikhitsa10,klinkhamer12,pa12,pazy13}.
For example, in the picture discussed in Ref. \cite{milgrom99} inertia (standing, more generally for the free matter actions) is determined and shaped by the interaction of matter with the quantum vacuum, which, in turn, is known to be affected by the cosmological state. Then, the $\az$ coincidence is natural, but the symmetry connection is not accounted for. For example, even in an exact de Sitter Universe one still gets a Newtonian behavior for high accelerations. However, the symmetry connection points to a DML dynamics in an exact de Sitter Universe.
\par
Another possibility is of an umbrella theory that accounts both for cosmology and local, galactic dynamics, in which the same theory constant enters both as a cosmological constant, and as $\az$ in local dynamics. This happens for example, in MOND versions of Einstein-Aether theories, in BIMOND, and in the theory of dipolar dark matter, all of which have a cosmological constant  appearing naturally. And, if we grant that these theories do not involve more than one acceleration (or length) constant, then the same constant appears both in the role of $\az$ in local dynamics, and as the cosmological constant. Here too, a symmetry connection between MOND and cosmology has not been identified.
\par
Besides looking for the desired, fundamental MOND theory, one can explore the cosmological implications of the known theories. Indeed, various aspects of cosmology and structure formation in some {\it subclasses} of known relativistic formulations of MOND were explored, e.g., in Refs. \cite{milgrom09,cz10,milgrom10b} (BIMOND), Refs. \cite{dodelson06,skordisetal06,skordis06,skordis08,zlosnik08} (TeVeS), and Ref. \cite{blanchet13} (dipolar dark matter). But these were far from exhaustive, and in any event, as said above, may have explored from the wrong starting points.
While, MOND does have elements that might replace the role of DM and dark energy, it has not been demonstrated that any of the known theories accounts for cosmology in all its observed details. \par
Thus, MOND as a paradigm (as contrasted with specific MOND formulations) does not yet make predictions regarding cosmology and structure formation, as it does for galactic systems.
\par
To me, the quest for a fundamental MOND theory
evokes in many ways the quest for a theory of quantum gravity. Note, e.g., the incapability of the standard dynamics to treat cosmology in the era when quantum effects are important -- along with other issues that require such a theory, e.g., quantum aspects of black holes -- because we lack such a theory . Here too, one tries to make various educated extrapolations of what we know about either gravity or quantum mechanics in regimes where they are known to work separately.
\par
Ideas proposed as part of one quest may also inspire ideas to proceed with the other, as was the case in Refs. \cite{rom10,blanmars11,sanders11}.
And, perhaps in the ultimate quantum-gravity theory $\hbar$ will also lose its meaning as a fundamental constant, and will turn out to be only an emergent effective constant relevant in weak-gravity quantum systems.

\section{\label{dmlem}MOND and nongravitational phenomena}
An important fact to note is that all our observational input and constraints on MOND come from systems that are controlled  by gravity. We thus do not have any experimental or observational hint on whether MOND is relevant for other phenomena such as EM. This obviously very interesting question has not received much attention, perhaps because of the lack of observational guidance, perhaps because of other, no less pressing questions.
\par
MOND might constitute a modification of gravity alone, and its effects on other (``matter'') degrees of freedom might enter only through their interaction with gravity. For example, bending of light \`a-la-MOND clearly enters EM.
However, as detailed above, there is a very promising possibility that MOND involves not only ``modification of gravity'', but constitutes a ``modification of inertia'', namely it entails modification of the free action of particles. But then it surely has to affect the actions of other matter degrees of freedom as well.
In this case MOND will clearly apply directly to nongravitational phenomena, i.e., it will show up, at some level, even when gravity can be neglected.
\par
But, beyond the change in dynamics ``at low accelerations'' that must apply to all degrees of freedom, there might be other doors through which $\az$ might enter other phenomena.
\par
Such extensions of MOND, if they apply at all, will have to be guided by theoretical considerations.
Here, I mention briefly some possible pointers; a more extensive discussion will be given elsewhere.
\par
What principle should we adopt for this task?
One possibility that I will follow here is simply to stick to the three basic tenets and construct modifications that a. use  $\az$, or the MOND length $\lm\equiv \az^{-1}$, b. require restoration of standard dynamics in the limit $\az\rar 0$ ($\lm\rar \infty$), and c. require SI of the field equations in the opposite limit, $\az\rar \infty$ ($\lm\rar 0$).
\par
Here we have to recall the discussion in Sec. \ref{structure} [below eq.(\ref{manare})] of how DML SI is achieved in the phenomenologically motivated, pure-gravity sector. We saw that it can be achieved by having $I\_G$ of a well defined scaling dimension $\xi$. Because we want DML SI of the whole theory, the choice of $\xi$ of the gravity sector dictates the same dimension for the DML of the nongravitational action.
\par
Consider, for example, quantum phenomena. The appearance of $\hbar$ is a block to SI as that of $G$ is in pure gravity (unless all masses vanish); it has units $[\hbar]=[\ell][m]$, so its value changes under scaling of the length-time units. Or, again, we can blame the masses, as is usually done, because we can work in units where $\hbar=1$, or accompany the scaling by inverse scaling of the mass units, so $\hbar$ is unchanged, but masses are not invariant.
If, however, we introduce $\az$ or $\lm$, we can have a theory where only the product $\hbar\az$, and possibly $c$, appear, all of which are invariant to unit scaling, so there is no obstacle to SI in this case.
\par
For example, take the Klein-Gordon action for a (real) massive scalar, whose action (in 4 dimensions) is \beq I_s\propto\int(\vfi\_{,\m}{\vfi\_,}\^\m+k\_m^2\vfi^2)d^4x, \eeqno{kgac} where $k\_m=m/\hbar$ is the Compton wave number associated with the scalar's mass. Having the benefit of $\lm$, we can generalize this action to (still keeping only first derivatives and separating the $\vfi$ and $\vfi\_{,\m}$ parts)
\beq I_s\propto \int[U(\lm^4\vfi\_{,\m}{\vfi\_,}\^\m)+(k\_m\lm)^2
V(\lm^2\vfi^2)]d^4x. \eeqno{polae}
Taking in the limit $\az\rar 0$, $U(y\rar\infty)\rar y,~V(z\rar\infty)\rar z$, we restore the action in eq.(\ref{kgac}).
Take, in the opposite, DML, $\az\rar\infty$, $U(y\rar 0)\rar y^\kappa,~V(z\rar 0)\rar z^\r$, the action becomes
  \beq I_s\propto  \int[(\vfi\_{,\m}{\vfi\_,}\^\m)^\kappa+(k\_m\lm)^2\lm^{2(\r-2\kappa)}
\vfi^{2\r}]d^4x. \eeqno{polsu}
If the scaling dimension of $\vfi$, which we can take at will, is $\xi$, then the scaling dimension of the first term is $2\kappa(\xi-1)+4$, and that of the second term is $2\r\xi+4$. SI of the equations of motion requires that these two dimensions are the same, and are equal to  that of other terms in the actions, e.g., of $I\_G$.
Thus if $I\_G$ is scale invariant, we want to have $\kappa(\xi-1)=\r\xi=-2$. For example, we could have $\kappa=1$, in which case the derivative term is not modified, which implies, $\xi=-1$, and thus $\r=2$, yielding the famously SI, $\vfi^4$ theory with the action
\beq I_s\propto\int(\vfi\_{,\m}{\vfi\_,}\^\m+\eta^{2}\vfi^4)d^4x, \eeqno{kiure}
where $\eta=k\_m\lm=m/\hbar\az$ is dimensionless.
\par
We saw that in MG MOND theories, $I\_G$ is not SI, but has scaling dimension $1$. To match this we need to have $\kappa(\xi-1)=\r\xi=-3/2$. Again, there are many solutions. For example, we may still leave the first term intact ($\kappa=1$), so $\xi=-1/2$, and $\r=3$, giving a $\vfi^6$ theory.
Generally, if we normalize the scalar field so that its scaling dimension matches its units dimension we must get an action in which only $\hbar\az$ appears. Indeed, defining $\psi\equiv \lm^{1+\xi}\vfi$
($\vfi$ having dimension $[\vfi]=[\ell]^{-1}$) we get
 \beq I_s\propto \int[(\psi\_{,\m}{\psi\_,}\^\m)^\kappa+(k\_m\lm)^2
\psi^{2\r}]d^4x. \eeqno{polmup}
\par
In the quantum context, the absolute value of the action is of material significance so to achieve SI we would want it to have it truly SI, i.e., have $\xi=0$.
\par
Consider now EM. The Maxwell action is SI; so if $I\_G$ is modified so that its DML is SI (i.e., has zero scaling dimension), we need not modify EM to get a proper DML of the full gravity-EM theory. However, if the scaling dimension of the DML of $I\_G$ is non-zero, we can modify Maxwell's theory to have a DML action with the same dimension.
The Maxwell action for a system of charges $e_i$ (in flat space-time) is
\beq I\_{EM}=-\frac{1}{16\pi}\int \fmn\Fmn d^4x
+\sum_i e_i\int \frac{dx\_i\^\m}{d\tau}A_\m[x_i(\tau)]d\tau_i\eeqno{vyut}
($\fmn=A_{\n,\m}-A_{\m,\n}$).
Indeed, with the vector potential, $A_\m$, having a scaling dimension $\xi=-1$, $I\_{EM}$ has $\xi=0$. As in the scalar case, there are various ways to write a modified DML for $I\_{EM}$ with other values of  $\xi$. For example, if we want $\xi=1$, we may give $A_\m$ a scaling dimension $\xi=-3/2$, then the first term in $I\_{EM}$ has dimension 1 and need not be modified. We may then modify the second term appropriately. In this case, the standard particle equation of motion $m_id^2x_i^\m/d\tau^2=e_iF_\n^\m(dx_i^\n/d \tau)$
will be modified.
\par
Taking another route, we can give $A_\m$ scaling dimension 0, so that the second term has $\xi=1$.
In this case the particle equation of motion is not modified, but becomes SI (both sides are multiplied by $\l^{-1}$ in a scaling transformation).
The first term has $\xi=2$ and needs to be modified to have a DML with $\xi=1$.\footnote{Such values of $\xi$ for $A_\m$ do not match its units dimensions (-1), but we can normalize $A_\m$ by a power of $\az$ to have a match. Because under scaling of the units $e\rar \lambda^{-1/2}e$, we would have in a SI theory charges always appearing as $e^2 a_0$. In the DML $\az\rar\infty,~e_i\rar 0$, such that $e_i^2 a_0$ remain finite.}
\par
On dimensional grounds, it is not possible to do this with $\lm$ alone. We need another constant that has also mass units. One possibility is to use $G$, which does not enter standard EM as such (or equivalently we can use $\azg$), to form a MOND scale of EM field
\beq E\_M\^G\equiv \az G^{-1/2}=\az^{3/2}\azg^{-1/2}\approx 4.6\times 10^{-5}{\rm cgs}. \eeqno{luksa}
Alternatively, we can use $\hbar$, which would bring in quantum aspects into ``classical'' EM, to form
\beq E\_M^h\equiv \left(\frac{\hbar}{\lm^4}\right)^{1/2}=
E\_M\^G\left(\frac{\ell\_p}{\lm}\right)
\approx 10^{-66}{\rm cgs}, \eeqno{lupla}
where $\ell\_p$ is the Planck length.
Or, if we could use some elementary charge, $e_0$, we could form
\beq E\_M^e\equiv\frac{e_0}{\lm^2}=\a_0^{1/2}E\_M^h, \eeqno{lumja}
 where $\a_0$ is the fine-structure constant associated with $e_0$.
Such constants do not otherwise appear in classical EM.
\par
At any rate, using such an electromagnetic-field constant, called hereafter $E\_M$, we can modify the Maxwell action in a way similar to the Born-Infeld modification, replacing it by\footnote{To retain gauge invariance we still keep the dependence on $A_\m$ through $\fmn$.}
\beq -\frac{\EM^2}{8\pi}\int\tilde\U(\fmn/\EM)d^4x. \eeqno{buta}
\par
Lorentz invariance dictates that the Lagrangian depends on $\fmn$ through its Lorentz invariants. In four dimensions these are
\beq P=\frac{1}{4}\fmn\Fmn=\oot(B^2-E^2),~~~~Q=\frac{1}{8}\eps^{\a\b\m\n}
F_{\a\b}\fmn=-\oot\vE\cdot\vB, \eeqno{lirom}
where $\vE$ and $\vB$ are the electric and magnetic fields ($\eps_{\a\b\m\n}$ is the totally antisymmetric tensor).
So write the EM field action as
\beq -\frac{\EM^2}{4\pi}\int\U(P/\EM^2,Q/\EM^2)d^4x. \eeqno{butama}
[In the standard Born-Infeld theory
$\U(a,b)\propto (1+2a-4b^2)^{1/2}$.]
\par
We want the theory to tend to Maxwell's in the limit of strong fields; so, in the limit $\EM\rar 0$, $\U(a,b)\rar a$.
In the opposite limit of $P,Q\ll \EM$, $\U$ has to have a scaling dimension as desired. For example, if we want $\xi=1$ for the modified Maxwell action, $\U$ has to have $\xi=-3$, while $P$ and $Q$ have $\xi=-2$; so  $\U$ becomes homogeneous of degree $3/2$, namely, in this limit $\U(\zeta a,\zeta b)=\zeta^{3/2}\U(a,b)$.


\begin{thebibliography}{}
\bibitem{milgrom83}M. Milgrom, Astrophys. J. 270, 365 (1983).
\bibitem{fm12}B. Famaey and S. McGaugh, Living Rev. Relativity 15, 10 (2012).
\bibitem{milgrom09a}M. Milgrom, Astrophys. J. 698, 1630 (2009).
\bibitem{milgrom14b}M. Milgrom, Phys. Rev. D 89, 024027  (2014).
\bibitem{milgrom14}M. Milgrom, Mon. Not. R. Astron. Soc. 437, 2531 (2014).
\bibitem{bm84}J. Bekenstein and M. Milgrom, Astrophys. J. 286, 7 (1984).
\bibitem{sanders90}R.H. Sanders,  Astron.
Astrophys. Rev., 2, 1(1990).
\bibitem{mcgaugh04}S. S. McGaugh, Astrophys. J. 609, 652 (2004).
\bibitem{scarpa06}R. Scarpa, AIP Conf. Proc. 822, 253, arXiv:astro-ph/0601478 (2006).
\bibitem{tiret09}O. Tiret and F. Combes, Astron. Astrophys. 496, 659 (2009).
\bibitem{milgrom09b}M. Milgrom, Mon. Not. R. Astron. Soc. 398, 1023 (2009).
\bibitem{trippe14}S. Trippe, Z. Naturforsch. 69a, 173 (2014).
\bibitem{wl14}M.G. Walker and A. Loeb, Contemp. Phys. 55, 198 (2014).
\bibitem{milgrom89}M. Milgrom, Comm. Astrophys. 13 (4), 215 (1989).
\bibitem{milgrom94}M. Milgrom, Ann. Phys. 229, 384 (1994).
\bibitem{milgrom99}M. Milgrom, Phys. Lett. A 253, 273 (1999).
\bibitem{brada95}R. Brada and M. Milgrom, Mon. Not. R. Astron. Soc. 276, 453 (1995).
\bibitem{angus12}G. Angus  et al., Mon. Not. R. Astron. Soc. 421, 2598 (2012).
\bibitem{milgrom12c}M. Milgrom, Phys. Rev. Lett. 109, 251103 (2012).
\bibitem{chang08}Z. Chang and X. Li, Phys. Lett. B 668, 453 (2008).
\bibitem{lichang12}X. Li and Z. Chang, arXiv:1204.2542 (2012).
\bibitem{deffayet11}C. Deffayet, G. Esposito-Farese, and R.P. Woodard, Phys. Rev. D 84, 124054 (2011).
\bibitem{milgrom14a}M. Milgrom, Phys. Rev. D 89, 024016  (2014).
\bibitem{milgrom97}M. Milgrom, Phys. Rev. E 56, 1148 (1997).
\bibitem{georgi07}H. Georgi, Phys. Rev. Lett. 98, 221601 (2007).
\bibitem{milgrom11}M. Milgrom, Acta Physica Polonica B vol. 42, 2175 (2011b).
\bibitem{izu}C. Itzykson and J.B. Zuber, {\it Quantum Field Theory}, McGraw-Hill (1980). \bibitem{milgrom10a}M. Milgrom,  Mon. Not. R. Astron. Soc. 403, 886 (2010).
\bibitem{bek04}J.D. Bekenstein, Phys. Rev. D 70, 083509 (2004).
\bibitem{sanders97}R.H. Sanders, Astrophys. J. 480, 492 (1997).
\bibitem{skordis09}C. Skordis, Class. Quant. Grav.
26, 143001 (2009).
\bibitem{fs09}P.G. Ferreira, and G.D. Starkman, Science 326, 812
(2009).
\bibitem{sagi9}E. Sagi, Phys. Rev. D 80, 044032 (2009).
\bibitem{freire12}P.C.C. Freire et al., Mon. Not. R. Astron. Soc. 423, 3328 (2012).
\bibitem{dodelson06}S. Dodelson
and M. Liguori, Phys. Rev. Lett. 97, 231301 (2006). \bibitem{skordisetal06}C. Skordis, D.F. Mota,
P.G. Ferreira, and C. Boehm, Phys. Rev. Lett. 96, 011301 (2006).
\bibitem{skordis06}C. Skordis, Phys. Rev. D 74,
103513 (2006).
\bibitem{skordis08}C. Skordis, Phys. Rev. D 77,
123502 (2008).
\bibitem{zlosnik08}T.G. Zlosnik, P.G. Ferreira, and G.D. Starkman,
Phys. Rev. D 77, 084010, (2008).
\bibitem{sagi10}E. Sagi, Phys. Rev. D 81, 064031 (2010).
\bibitem{babichev11}E. Babichev,  et al., Phys. Rev. 84, 061502 (2011). k-mouflage
\bibitem{jm01}T. Jacobson and D. Mattingly, Phys. Rev. D 64,
024028 (2001).
\bibitem{zlosnik07}T.G. Zlosnik, P.G. Ferreira, and G.D. Starkman, Phys. Rev. D 75 044017 (2007).
\bibitem{milgrom09}M. Milgrom, Phys. Rev. D 80, 123536 (2009).
\bibitem{milgrom10b}M. Milgrom, Mon. Not. R. Astron. Soc. 405, 1129 (2010).
\bibitem{milgrom10c}M. Milgrom, Phys. Rev. D 82, 043523 (2010).
\bibitem{cz10}T. Clifton and T.G. Zlosnik, Phys. Rev. 81, 103525 (2010).
\bibitem{soussa03}M.E. Soussa and R.P. Woodard, Class. Quant. Grav. 20, 2737 (2003).
\bibitem{blanchet07}L. Blanchet, Class. Quant. Grav. 24, 3541 (2007).
\bibitem{blt08}L. Blanchet, and A.  Le Tiec,    Phys. Rev. D 78, 024031 (2008).
\bibitem{blt09}L. Blanchet, and A.  Le Tiec,
Phys. Rev. D 80, 023524 (2009).
\bibitem{pikhitsa10}P.V. Pikhitsa, arXiv1010.0318 (2010).
\bibitem{kt10}V. V. Kiselev and S. A. Timofeev, arXiv:1009.1301
(2010).
\bibitem{lc11}X. Li and Z. Chang, Commun. Theor. Phys. 55, 733 (2011).
\bibitem{klinkhamer12}F.R. Klinkhamer, Mod. Phys. Lett. A 27, 1250056 (2012).
\bibitem{pa12}E. Pazy and N. Argaman, Phys. Rev. D 85, 104021 (2012).
\bibitem{pazy13}E. Pazy, Phys. Rev. D 87, 084063 (2013).
\bibitem{milgrom02}M. Milgrom, New Astron.Rev. 46, 741 (2002).
\bibitem{bruneton09}J-P. Bruneton, S.
Liberati, L. Sindoni, and B. Famaey, JCAP, 03, 021 (2009).
\bibitem{zl10}H. Zhao and B. Li,  Astrophys. J. 712, 130 (2010).
\bibitem{ho10}C.M. Ho, D. Minic, Y.J. Ng, Phys. Lett. B693, 567
(2010).
\bibitem{ho12}C.M. Ho  et al., Phys. Rev. D 85, 104033 (2012).
\bibitem{bettoni11}D. Bettoni, S. Liberati, and L. Sindoni, J. Cosm. Astropart. Phys. 11, 007 (2011).
\bibitem{rom10}J.M. Romero, R. Bernal-Jaquez, O.
Gonz\'{a}lez-Gaxiola, Mod. Phys. Lett. A, 25 (29), 2501 (2010).
\bibitem{blanmars11}L. Blanchet and S. Marsat, Phys. Rev. D 84, 044056 (2011).
\bibitem{sanders11}R.H. Sanders, Phys. Rev. D 84, 084024 (2011).
\bibitem{bernal11}T. Bernal, S. Capozziello, J.C. Hidalgo, and S. Mendoza, Europ. Phys. J. C 71, 1794 (2011).
\bibitem{trippe13}S. Trippe, J. Korean Astron. Soc. 46, 93 (2013).
\bibitem{bl68}H. Bacry and J.M. L\'evy-Leblond, J. Math. Phys., 9, 1605 (1968).
\bibitem{dyson72}F.J. Dyson, Bull. Am. Math. Soc. 78 (5), 635 (1972).
\bibitem{aldrovandi09}R. Aldrovandi and J.P. Pereira, Int. J. Mod. Phys. D17, 2485 (2009).
\bibitem{blanchet13}L. Blanchet, D. Langlois, A. Le Tiec, and S. Marsat, J. Cosm. Astropart. Phys. 22, 1302 (2013).


\end{thebibliography}
\end{document}